\documentclass[11pt, a4paper]{article}
\pdfoutput=1

\usepackage{amsmath}
\usepackage{amsfonts}
\usepackage{amssymb}
\usepackage{bm}
\usepackage{bbm}
\usepackage{verbatim}
\usepackage{dsfont}
\usepackage{booktabs}
\usepackage{slashed}
\usepackage{nicefrac}
\usepackage{mathtools}

\usepackage{epsfig}
\usepackage{color}
\usepackage[table]{xcolor}
\usepackage{graphicx}
\usepackage[font=small]{caption}
\usepackage[listofformat=empty,subrefformat=empty]{subfig}

\usepackage{float}
\usepackage{overpic}
\usepackage{tikz}

\usepackage{enumerate}
\usepackage{hhline}
\usepackage{multirow}

\usepackage[nosort]{cite}
\usepackage{xspace}
\usepackage{setspace}

\usepackage{youngtab}

\usepackage[pdftitle={},
  pdfauthor={},
  pdfsubject={},
  bookmarksopen, bookmarksnumbered, bookmarksopenlevel=2, colorlinks=false, linkcolor=blue, citecolor=blue, urlcolor=blue]{hyperref}

\usepackage[nameinlink,capitalize]{cleveref}

\makeatletter{}
\g@addto@macro\bfseries{\boldmath}
\makeatother

\DeclareMathOperator{\tr}{tr}

\begin{document}

% format
\baselineskip=18pt  % a la harvmac
\numberwithin{equation}{section}  % make eq labels (sec.num)
\allowdisplaybreaks  % allow page breaks in displayed eqs

%%%%%%%%%%%%%%%%%%%%%%%%%%%%%%%%%%%%%%%%%%%
%%%        TITLE BEGINS HERE
%%%%%%%%%%%%%%%%%%%%%%%%%%%%%%%%%%%%%%%%%%%

\thispagestyle{empty}

\vspace*{-2cm}
% \begin{flushright}
% 	\texttt{report number}
% \end{flushright}

% title, authors, affiliation
\vspace*{1.6cm}
\begin{center}
{\LARGE{\textbf{Anomaly Inflow and Gauge Group Topology\\ \vspace{.4cm}
in the 10d Sugimoto String Theory
}}} \\
\vspace*{1.5cm}
Vittorio Larotonda and Ling Lin\\

\vspace*{1.0cm}
{\it Dipartimento di Fisica e Astronomia, Universit\`{a} di Bologna, via Irnerio 46, Bologna, Italy,}\\
\vspace*{.1cm}
{\it INFN, Sezione di Bologna, viale Berti Pichat 6/2, Bologna, Italy}

\vspace*{0.8cm}
\end{center}
\vspace*{.5cm}

\noindent
We revisit the chiral spectra on charged 1- and 5-branes in the 10d non-supersymmetric $\mathfrak{sp}(16)$ string theory (also known as the Sugimoto model), and verify that they consistently cancel the anomaly inflow induced by a Green--Schwarz mechanism in the bulk.
By analyzing the $\mathfrak{sp}(16)$ representations arising from quantizing the fermion zero modes on these branes as well as uncharged 4-branes, we find compelling evidence that the global structure of the gauge group is $Sp(16)/\mathbb{Z}_2$.
We further comment on a possible duality to non-supersymmetric heterotic strings, and explore bottom-up anomaly inflow constraints for 10d effective $Sp(16)/\mathbb{Z}_2$ gauge theories coupled to gravity.

\newpage
%%%%%%%%%%%%%%%%%%%%%%%%%%%%%%%%%%%%%%%%%%%
%%%           TITLE ENDS HERE
%%%%%%%%%%%%%%%%%%%%%%%%%%%%%%%%%%%%%%%%%%%
\setcounter{tocdepth}{2}
\tableofcontents
%\printindex

% \newpage

%%%%%%%%%%%%%%%%%%%%%%%%%%%%%%%%%%%%%%%%%%%
%%%        MAIN TEXT BEGINS HERE
%%%%%%%%%%%%%%%%%%%%%%%%%%%%%%%%%%%%%%%%%%%

\section{Introduction}

Non-supersymmetric string theories provide a rich framework for exploring quantum gravity and its consistency constraints in scenarios where supersymmetry is absent, and have seen a resurgence of interest, see, e.g., \cite{Abel:2016pwa, Abel:2018zyt, McGuigan:2019gdb, Itoyama:2020ifw, Angelantonj:2020pyr, Basile:2020mpt, Basile:2020xwi, Kaidi:2020jla, Faraggi:2020hpy, Cribiori:2020sct, Itoyama:2021fwc, Gonzalo:2021fma, Perez-Martinez:2021zjj, Basile:2021mkd, Basile:2021vxh, Sagnotti:2021mxb, Itoyama:2021itj, Cribiori:2021txm, Raucci:2022bjw, Basile:2022ypo, Baykara:2022cwj, Koga:2022qch, Cervantes:2023wti, Avalos:2023mti, BoyleSmith:2023xkd, Raucci:2023xgx, Mourad:2023wjg, Avalos:2023ldc, Fraiman:2023cpa, Basile:2023knk, DeFreitas:2024ztt, Tachikawa:2024ucm, Saxena:2024eil, Baykara:2024tjr, Baykara:2024vss, Angelantonj:2024jtu, DeFreitas:2024yzr, Basaad:2024lno, Raucci:2024fnp, Abel:2024vov}.
In light of these efforts, a deeper and more thorough understanding of these string theories is clearly desirable.

In this work we focus on the 10d Sugimoto model \cite{Sugimoto:1999tx}, which can be constructed from type IIB string theory with a spacetime filling orientifold plane with positive tension and Ramond-Ramonod charge (an O9$^+$).
This setup which bears close resemblance to type I superstring requires 32 9-branes to cancel tadpoles.
The notable differences to type I is that the positive RR-charge of the O9$^+$ requires \emph{anti-}D9s, which breaks spacetime supersymmetry in the open string sector.
Moreover, the different orientifold action leads to a spacetime $\mathfrak{usp}(32) \equiv \mathfrak{sp}(16)$ gauge symmetry, as opposed to $\mathfrak{so}(32)$ in type I.
Another similarity to type I is the anomalous chiral fermion spectrum, which necessitates a Green--Schwarz (GS) mechanism.
Our goal then will be to show that a consistent anomaly cancellation in this fashion also implies the gauge \emph{group} to be $Sp(16)/\mathbb{Z}_2$.

The main players that tie together these two aspects are the 1- and 5-branes that couple electromagnetically to the chiral 2-form gauge field $B_2$ required by the Green--Schwarz mechanism.\footnote{These defects are inherited from the D1-/D5-branes of type IIB, and are typically also referred as such in the literature.
However, the objects that survive the orientifolds, both in type I and the Sugimoto model, may be more appropriately described by bound states of two D-branes \cite{Gimon:1996rq,Sugimoto:1999tx}.
To emphasize this possible difference between pre- and post-orientifold branes, we use $p$-branes rather than D$p$-branes for the latter.
}
Namely, the GS-cancellation of the 10d chiral anomaly, captured by the factorizable anomaly polynomial $I_{12} = X_4 \wedge X_8$, induces an anomaly inflow onto the branes.
Being dynamical defects in the 10d bulk themselves, quantum consistency of the overall system requires the anomaly of the worldvolume degrees of freedom, arising from open strings ending on the defects, to cancel this inflow \cite{Callan:1984sa}: $I_4 = X_4$ for a 1-brane and $I_8 = X_8$ for a 5-brane.
While this condition is long-known to be met by the charged defects in type I \cite{Dixon:1992if,Mourad:1997uc}, it has not been explicitly shown for the Sugimoto model before.
While completely analogous to type I, our analysis in this paper shows that, due to the exchanged $\mathfrak{so}$- and $\mathfrak{sp}$-Chan--Paton factors on the 1- and 5-branes, a consistent inflow in the Sugimoto model requires different normalizations of $X_{4/8}$ compared to type I.\footnote{This comparison is sensible because the 10d anomaly polynomial of type I and the Sugimoto model are identical upon exchanging $\mathfrak{so}(32)$ and $\mathfrak{sp}(16)$ characteristic classes.}
This leads, in particular, to different 6d Green--Schwarz terms on the 5-branes of type I and the Sugimoto model.

The same localized degrees of freedom that facilitate the consistent inflow matching also provide a non-perturbative access to the gauge group topology of the 10d effective theory.
This information is relevant, e.g., to understanding string dualities and the structure of the moduli space after compactification, as is evident from analogous results in type I \cite{Witten:1997bs}.
However, the perturbative spectrum (perturbative from a spacetime effective gauge theory perspective) of both type I and the Sugimoto model contains only massless fields in center-invariant representations and thus cannot distinguish between different global forms of the spacetime gauge group.
On the other hand, the spectrum on the branes constrained by inflow may give rise to spacetime excitations that transform non-trivially under the center.
The prime example that inspire our subsequent study is the 0-brane in type I, which are stable non-BPS particles transforming the spinor representation of $\mathfrak{so}(32)$ \cite{Sen:1998tt,Sen:1998ki,Sen:1999mg}, and were needed to corroborate the type I / heterotic duality.
To see the spinor representation, one must quantize the open string modes on the 0-brane worldline \cite{Witten:2023snr}, whose analogues on the 1- and 5-branes are precisely the chiral fields supplying the anomaly.

In the absence of stable 0-branes in the Sugimoto model \cite{Sugimoto:1999tx}, it was speculated \cite{Basile:2023knk} that the gauge group is $Sp(16)/\mathbb{Z}_2$.
By explicitly quantizing the fermion zero modes of the Green--Schwarz defects (i.e., the 1- and 5-branes) as well as the uncharged stable 4-brane along the lines of \cite{Witten:2023snr}, we confirm the absence of center-non-invariant representations, thus providing strong support for the non-simply-connected gauge group topology.
We also point out a potential ``S-like duality'' of the Sugimoto model to non-supersymmetric heterotic strings based on this gauge group.

Motivated by the important role the Green--Schwarz defects play in the consistency of the Sugimoto model, we also perform a bottom-up study of anomaly-inflows of effective 10d $\mathfrak{sp}(16)$ gauge theories coupled to gravity.
In this analysis, we allow for the spacetime theory to have varying massless chiral spectra compatible with a $Sp(16)/\mathbb{Z}_2$ group, require chiral anomalies to be cancelled by a GS-mechanism.
Allowing also for a varying chiral spectrum on the associated 1- and 5-branes (though fixed to have the same gauge algebra as in the Sugimoto model), we find that matching the inflow is strong enough to eliminate all possibilities but the one matching the Sugimoto model within the classes of models we consider.
In the absence of any supersymmetry that can completely fix the massless spectrum, this serves as evidence that the Sugimoto model may be the only consistent realization of a 10d effective $Sp(16)/\mathbb{Z}_2$ gauge theory coupled to gravity.

The rest of the paper is structured as follows.
In Section \ref{sec:anomaly_inflow} we explicitly verify a consistent anomaly inflow onto 1- and 5-branes required to cancel the 10d chiral anomaly via a Green--Schwarz mechanism, and point out the similarities and differences between type I and Sugimoto.
The required worldvolume fermion spectra is then, in Section \ref{sec:defect_spectra}, shown to give rise to spacetime excitations compatible with gauge groups $Spin(32)/\mathbb{Z}_2$ and $Sp(16)/\mathbb{Z}_2$, respectively.
In these sections we perform the analysis of the Sugimoto model, which leads to novel results, alongside an analysis of type I, given their many similarities.
In Section \ref{sec:bottom-up}, we instead focus solely on the bottom-up study of anomaly matching constraints with $\mathfrak{sp}(16)$ gauge symmetry.

\section{Anomaly inflow in orientifolds of IIB}
\label{sec:anomaly_inflow}

In this section we will revisit the Green--Schwarz mechanism in type I and the Sugimoto model and the associated anomaly inflow.

\subsection{Chiral anomalies and the Green--Schwarz mechanism}

Despite the major difference in the amount of supersymmetry, the two orientifolds of type IIB have similar gauge sectors and, in fact, identical anomaly polynomials.
While the gauge algebras of type I and the Sugimoto model are $\mathfrak{so}(32)$ and $\mathfrak{usp}(32) \equiv \mathfrak{sp}(16)$, respectively, the fermion spectrum in both cases can be summarized as follows:
\begin{align}
\renewcommand{\arraystretch}{1.3}
    \begin{array}{c|c|c}
        \text{fields} & \mathfrak{g}_B \text{-rep} & \text{chirality} \\ \hline
        \phi \, (\text{``gaugino''}) & \Yvcentermath1\tiny\yng(1,1) & + \\
        \psi \, (\text{gravitino}) & {\bf 1} & + \\
        \lambda \, (\text{dilatino}) & {\bf 1} & -  
    \end{array}
\end{align}
While for $\mathfrak{g}_B = \mathfrak{so}(32)$ the anti-symmetric representation (denoted using the usual Young tableau notation) is the adjoint ${\bf 496}$, it decomposes into $\Yvcentermath1\tiny\yng(1,1) = {\bf 495} \oplus {\bf 1}$ for $\mathfrak{g}_B = \mathfrak{sp}(16)$.\footnote{For $\mathfrak{sp}$ the adjoint representation is the symmetric tensor product of the fundamental; $\phi$ can therefore not be the superpartner of the gauge bosons, consistent with supersymmetry breaking in the Sugimoto model.
}
The singlet is the goldstino associated with supersymmetry breaking in the open string sector of the Sugimoto model.
The closed-string sectors are identical, and give rise to the gravitino and dilatino (retaining the names also in the Sugimoto model despite broken supersymmetry).

These fermions contribute to the chiral anomaly of the theories.
The individual fields have the standard anomaly polynomial of a Weyl fermion in $d$ dimensions \cite{Alvarez-Gaume:1983ihn},
\begin{align}
    I_{d+2}^{\chi} = \pm\frac{1}{2} \left[ \hat{A}(R) \, \text{ch}_{\rho}(F) \right] \Big|_{d+2} \, ,
    \label{eq:Anomaly polynomial}
\end{align}
where $\hat{A}(R)$ is the A-roof genus of the Riemann curvature 2-form $R$, $\text{ch}_{\rho}(F)$ is the Chern character of the $\mathfrak{so}(32)$/$\mathfrak{sp}(16)$ field strength 2-form $F$ in the representation $\rho$ of the field $\chi$; the overall sign depends on the chirality and the factor $1/2$ is due to the fact that all fermions under consideration are Majorana--Weyl.
See Appendix \ref{sec:Appendix A} for more details and conventions.
The full anomaly polynomial for the 10d chiral anomaly is therefore
\begin{align}\label{eq:10d-anomaly-polynomial}
    \begin{split}
        I_{12}^\text{fermions} & = I_{12}^\psi + I_{12}^\lambda + I^\phi_{12} \\ 
        & = \frac{1}{4 \times 192}\left(\tr R^2 -\tr F^2\right) \left(8 \tr F^4-\tr F^2 \tr R^2+\tr R^4+\frac{1}{4} \left(\tr R^2\right)^2\right) \\
        & \equiv X_4 \wedge X_8 \, ,
    \end{split}
\end{align}
with
\begin{align}\label{eq:10d_factors_no-norm}
\begin{split}
    X_4 & = \frac{c}{4} (\tr R^2 - \tr F^2) = c \left(- \frac{p_1(R)}{2} - \frac{\tr F^2}{4} \right)  \, , \\
    X_8 & = \frac{1}{192c} \left(8 \tr F^4-\tr F^2 \tr R^2+\tr R^4+\frac{1}{4} \left(\tr R^2\right)^2 \right) \\
    & = \frac{1}{c} \left( \frac{3 p_1^2(R) - 4 p_2(R)}{192} + \frac{p_1(R) \, \tr F^2}{96} + \frac{\tr F^4}{24} \right)\, .
\end{split}
\end{align}
The normalizations here are intentionally left unspecified, as they will differ between type I and the Sugimoto model.

The fact that $I_{12} = X_4 \wedge X_8$ factorizes makes a cancellation of the anomaly via a Green--Schwarz (GS) mechanism possible.
For this, there needs to exist a dynamical 2-form $U(1)$ gauge field $B_2$ which transforms under the anomalous gauge transformations as $\delta B_2 = - X_2$, which satisfies the descent conditions $X_4 = d X_3$, $\delta X_3 = d X_2$.
Then, by adding the coupling 
\begin{align}\label{eq:GS-term}
    S_\text{GS} = \int B_2 \wedge X_8    
\end{align}
to the action, the descent procedure leads to the anomaly
\begin{align}
    \delta S_\text{GS} =  - \int X_2 \wedge X_8 \xrightarrow{\text{descent}} I_{12}^\text{GS} =  - \int X_4 \wedge X_8 \, ,
\end{align}
which cancels the chiral anomalies from the fermions, $I^\text{fermions}_{12} + I_{12}^\text{GS} =0$.
In both type I and the Sugimoto model, the closed string RR 2-form $B_2$ of IIB which survives the orientifold precisely fills this role.
Note that the descent equation is equivalent to the Bianchi identity, $d H_3 = X_4$, where $H_3 = d B_2 - \omega_\text{CS}$ is the gauge-invariant combination of the RR field strength and the Chern--Simons 3-form.

A crucial consistency condition is matching the spacetime anomalies via inflow with the anomalies localized on the defects that are charged under the 2-form participating in the Green--Schwarz mechanism.
These defects are the strings (1-branes) and 5-branes that the orientifolded models inherit from D1-/D5-branes of type IIB.
In the following, we shall verify that the inflow indeed matches the anomalies of the worldvolume theories, which hinges on the correct normalization in \eqref{eq:10d_factors_no-norm}.
Though for the 5-brane the computations are analogous to inflow in type 0 and heterotic string theories \cite{Mourad:1997uc, Dudas:2000sn}, it is, to our knowledge, the first explicit demonstration for the Sugimoto model.

\subsection{Anomaly inflow onto charged defects}

Anomaly inflow is a general feature when defects carry (electric or magnetic) charge $q$ under a $p$-form $B_p$ which is non-invariant under gauge symmetry transformations.
For a continuous gauge transformation $\delta B_p = - X_p$, which is the relevant type for us, the coupling $q \int_{\Sigma_p} B_p$ induces an anomaly on the worldvolume $\Sigma_p$ of the defect, which is characterized by the anomaly polynomial $I_{p+2}^\text{inflow} = - q X_{p+2}|_{\Sigma_p}$, which is related to $X_p$ via descent.
Consistency then requires that this inflow is cancelled by the worldvolume dynamics with anomaly polynomial $I_{p+2}$, $I^\text{inflow}_{p+2} = - I_{p+2}$.

The effective gauge theories that capture the worldvolume anomalies on the 1- and 5-branes depend on the type of orientifold involution which acts on IIB.
For type I, we have
\begin{itemize}
    \item an $\mathfrak{g}_W = \mathfrak{so}(n)$ gauge theory on a stack of $n$ 1-branes \cite{Polchinski:1995df, Gava:1998sv};
    \item an $\mathfrak{g}_W = \mathfrak{usp}(2n) = \mathfrak{sp}(n)$ gauge theory on a stack of $n$ 5-branes \cite{Witten:1995gx,Gimon:1996rq}.
\end{itemize}
Changing the orientifold projection effectively swaps the types of gauge algebras.
In the Sugimoto model, we thus have \cite{Sugimoto:1999tx}
\begin{itemize}
    \item an $\mathfrak{g}_W = \mathfrak{sp}(n)$ gauge theory on a stack of $n$ 1-branes;
    \item an $\mathfrak{g}_W = \mathfrak{so}(n)$ gauge theory on a stack of $n$ 5-branes.
\end{itemize}
Let us now turn to the spectra of these theories and compute the anomaly inflow.\footnote{Note that in type I, the 5-brane stack with $\mathfrak{g}_W = \mathfrak{sp}(n)$ is really a stack of $2n$ type IIB D5-branes (or, $n$ pairs of D5 and image-D5), which due to the orientifold move in pairs \cite{Gimon:1996rq}.
Applying the logic to the orientifold action for the Sugimoto model ``pairs'' D1s instead of D5s \cite{Sugimoto:1999tx}.
At the same time, the orientifold also ``halves'' the Dirac pairing between the D1s and D5s of IIB, so that ultimately the fundamental electric and magnetic charges carry the listed gauge algebras with $n=1$ in both models.
This may be seen as a top-down derivation of $X_4^\text{Sugimoto} = 2 X_4^\text{type I}$ and $X_8^\text{type I} = 2 X_8^\text{Sugimoto}$, which we now verify via inflow.
\label{footnote:half-D5s}
}

\subsubsection*{1-branes}

Apart from the difference in the gauge algebra $\mathfrak{g}_W$ on the strings between type I and the Sugimoto model, the two theories can be characterized in an identical way.
The systems global symmetries are the gauge algebra $\mathfrak{g}_B$ of the bulk theory, and the transverse rotations $\mathfrak{so}(8)_T$.
The chiral spectrum is analogous between the two models.
It contains three Majorana--Weyl fermions which have the following representations under the gauge and global symmetry algebras \cite{Polchinski:1995df,Sugimoto:1999tx}:
\begin{align}\label{tab:D1 spectrum}
\renewcommand{\arraystretch}{1.3}
    \begin{array}{c | c | c| c| c}
         & \text{chirality} & \mathfrak{so}(8)_T & \mathfrak{g}_B & \mathfrak{g}_W \\ \hline
         \chi^{(1)} & + & {\bf 8}_c & {\bf 1} & \Yvcentermath1\tiny\yng(1,1) \\
         \chi^{(2)} & - & {\bf 8}_s & {\bf 1} & \Yvcentermath1\tiny\yng(2) \\
         \chi^{(3)} & + & {\bf 1} & {\bf 32} & \Yvcentermath1\tiny\yng(1)
    \end{array}
\end{align}
Note that for $\mathfrak{g}_W = \mathfrak{so}(n)$, the symmetric representation is reducible and decomposes into the trace singlet and the symmetric traceless part, while for $\mathfrak{g}_W = \mathfrak{sp}(n)$ the anti-symmetric representation is reducible, as mentioned above.

The anomaly polynomial of the worldsheet theory is then the sum of the individual contributions of these fields,
\begin{align}
    I_4 = I_4^{\chi^{(1)}} + I_4^{\chi^{(2)}} + I_4^{\chi^{(3)}} \, ,
\end{align}
with
\begin{equation}
    I_4^{\chi} =\pm \frac{1}{2}\big[ \hat{A}(\tilde{R}) \, \text{ch}_r(N) \, \text{ch}_{{\cal R}_B} (F_B) \, \text{ch}_{{\cal R}_W} (F_W) \big] \big|_{4} \, .
\end{equation}
Compared to \eqref{eq:Anomaly polynomial}, the first two factors arise from the decomposition of the 10d tangent bundle ${\cal T}M_{10}|_{\Sigma_2} = {\cal T}\Sigma_2 \oplus {\cal N}\Sigma_2$ into the tangent and normal bundle of the string worldsheet $\Sigma_2$, with $\tilde{R}$ being the curvature of ${\cal T}\Sigma_2$ and $N$ the curvature of the normal bundle ${\cal N}\Sigma_2$ (associated with $\mathfrak{so}(8)_T$).
The field strength of the bulk / worldsheet gauge symmetry is $F_B$ / $F_W$.
Finally, $r$ is the representation under $\mathfrak{so}(8)_T$, and ${\cal R}_{B/W}$ is the representation under $\mathfrak{g}_B$ / $\mathfrak{g}_W$.

The computation then proceeds straightforwardly.
Following the conventions of appendix \ref{sec:Appendix A}, we find,
\begin{equation}
    \begin{split}
        I_4^{\chi^{(1)}} & = \frac{\dim( \, \Yvcentermath1\tiny\yng(1,1) \,)}{2} \left(p_1(N) -\frac{p_1(\tilde
        {R})}{3}\right) - 2 \tr_{ \, \Yvcentermath1\tiny\yng(1,1)} F_W^2\, , \\
        I_4^{\chi^{(2)}} & = - \frac{\dim( \, \Yvcentermath1\tiny\yng(2) \,)}{2} \left(p_1(N) -\frac{p_1(\tilde
        {R})}{3}\right) + 2 \tr_{\, \Yvcentermath1\tiny\yng(2)} F_W^2\, , \\
        I_4^{\chi^{(3)}} & = - \frac{2 \dim ( \, \Yvcentermath1\tiny\yng(1) \, )}{3}\, p_1(\tilde{R}) - \frac{\dim ( \, \Yvcentermath1\tiny\yng(1) \, )}{4}n\, \tr F_B^2 - 8 \tr F_W^2\, , 
    \end{split}
\end{equation}
where the traces without subscript are in the fundamental (vector) representation of $\mathfrak{g}_W$.
The dimensionality and trace relations take identical expressions for $\mathfrak{so}(k)$ and $\mathfrak{usp}(k) = \mathfrak{sp}(k/2)$:
\begin{align}
    \begin{split}
        \dim ( \, \Yvcentermath1\tiny\yng(1) \, ) & = k \, , \quad \dim ( \, \Yvcentermath1\tiny\yng(1,1) \, ) = \frac{k(k-1)}{2} \, , \quad \dim ( \, \Yvcentermath1\tiny\yng(2) \, ) = \frac{k(k+1)}{2} \, , \\
        \tr_{ \, \Yvcentermath1\tiny\yng(1,1)} F_W^2 & = (k-2) \tr F_W^2 \, , \quad \tr_{ \, \Yvcentermath1\tiny\yng(2)} F_W^2  = (k+2) \tr F_W^2 \, .
    \end{split}
\end{align}
Using these, we find that the total anomaly is
\begin{equation}
    I_4 = \sum_{i=1}^3 I_4^{\chi^{(i)}} = \begin{cases}
        -\frac{n}{2} (p_1(\tilde{R}) + p_1(N) ) - \frac{n}{4} \tr F_W^2 \, , & \text{type I } \, (\mathfrak{g}_W = \mathfrak{so}(n)) \, , \\
        - n (p_1(\tilde{R}) + p_1(N) ) - \frac{n}{2} \tr F_W^2 \, , & \text{Sugimoto } \, (\mathfrak{g}_W = \mathfrak{sp}(n)) \, .
    \end{cases}
\end{equation}
Considering that along the worldsheet $\Sigma_2$, the decomposition of the 10d tangent bundle leads to $\frac{1}{2} \tr R^2 = - p_1(R) = - p_1(\tilde{R}) - p_1(N)$, we thus find the following anomaly inflow onto a stacks of $n$ 1-branes:
\begin{align}\label{eq:inflow_D1s}
    n X_4|_{\Sigma_2} \stackrel{!}{=} - I_4^\text{inflow} = I_4 =   \begin{cases}
            \frac{n}{4}  (\tr R^2 - \tr F_W^2)|_{\Sigma_2} \, , & \text{type I} \, , \\
            \frac{n}{2} (\tr R^2 - \tr F_W^2)|_{\Sigma_2} \, , & \text{Sugimoto} \, .
    \end{cases}
\end{align}
This fixes the normalizations in \eqref{eq:10d_factors_no-norm} to be $c = 1$ for type I and $c = 2$ for Sugimoto.
This subtle difference between type I and the Sugimoto model, which to our knowledge has not been discussed in the literature, is tied to the different IIB orientifold projections and their action on open strings, see also footnote \ref{footnote:half-D5s}.

\subsubsection*{5-branes}

The inflow onto the magnetically charged 5-branes with worldvolume $\Sigma_6$ can be equivalently described by the coupling
\begin{equation}
    \int_{\Sigma_6} B_6
    \label{eq:D5 brane coupling}
\end{equation}
to the gauge potential that is magnetically dual to $B_2$.
In the presence of the Green--Schwarz term \eqref{eq:GS-term}, the equations of motion for $B_2$, $d H_7 = X_8$ with $H_7 = * H_3$, then implies that $\delta B_6 = -X_6$ which descents to $X_8$. 
With \eqref{eq:10d-anomaly-polynomial} and \eqref{eq:inflow_D1s}, the normalization of the 8-form is fixed, and must be matched by the anomaly on 5-brane stacks.

The chiral fermions of the effective gauge theory description on a stack of $n$ 5-branes are symplectic Majorana--Weyl fermions in the following representations \cite{ Polchinski:1995df, Witten:1995gx,Gimon:1996rq, Mourad:1997uc}:
\begin{align}\label{tab:D5 spectrum}
\renewcommand{\arraystretch}{1.3}
    \begin{array}{c|c|c|c|c}
        & \text{chirality} & \mathfrak{so}(4)_T & \mathfrak{g}_B & \mathfrak{g}_W \\ \hline
        \varphi^{(1)} & + & {\bf 2}_+ & {\bf 1} & \Yvcentermath1\tiny\yng(1,1) \\
        \varphi^{(2)} & - & {\bf 2}_- & {\bf 1} & \tiny\yng(2) \\
        \varphi^{(3)} & + & {\bf 1} & {\bf 32} & \tiny\yng(1)
    \end{array}
\end{align}
Just as for the Majorana--Weyl fermions on the 1-branes, a symplectic Majorana--Weyl contributes the anomaly 8-form $I^{\varphi}_8 = \pm\frac12 \big[ \hat{A}(\tilde{R}) \, \text{ch}_r(N) \, \text{ch}_{{\cal R}_B}(F_B) \, \text{ch}_{{\cal R}_W}(F_W) \big] \big|_8$.
A slightly more tedious computation, using the results in Appendix \ref{sec:Appendix A}, leads to the following anomaly polynomials:
\begin{align}
    \begin{split}
        I_8^{\varphi^{(1)}}  =\, & \frac{k(k-1)}{4}
        \Bigg[ \frac{7 p_1^2(\tilde{R}) - 4 p_2(\tilde{R})}{2880} + \frac{4 p_2(N) + p_1^2(N) - 2p_1(\tilde{R}) p_1(N)}{192} \\
        &+ \frac{\chi(N)[p_1(N) - p_1(\tilde{R})]}{48} \Bigg] +\frac{k-2}{48}\left(p_1(\tilde{R}) - 3p_1(N)-6\chi(N) \right)\tr F_W^2 \\ 
        & +\frac{1}{24}\left[3(\tr F_W^2)^2 +(k-8)\tr F_W^4\right] \, ,
    \end{split} \\
    \begin{split}
        I_8^{\varphi^{(2)}} = \, & \frac{k(k+1)}{4}
        \Bigg[ \frac{4p_2(\tilde{R}) - 7 p_1^2(\tilde{R})}{2880} - \frac{4 p_2(N) + p_1^2(N) - 2p_1(\tilde{R}) p_1(N)}{192}  \\
        & + \frac{\chi(N) [p_1(N) - p_1(\tilde{R})]}{48} \Bigg]- \frac{k+2}{48} \left(p_1(\tilde{R}) - 3p_1(N) + 6 \chi(N) \right) \tr F_W^2 \\
        & - \frac{1}{24}\left[3(\tr F_W^2)^2 +(k+8)\tr F_W^4\right] \, ,
    \end{split}\\
    \begin{split}
        I_8^{\varphi^{(3)}} = \, & \frac{k}{1440} \left[ 28 p_1(\tilde{R})^2 - 16 p_2(\tilde{R}) + 15 p_1(\tilde{R}) \tr F_B^2 + 30 \tr F_B^4 \right] \\
        & + \tr F_W^2 \left( \frac{\tr F_B^2}{8} + \frac{p_1(\tilde{R})}{3} \right) + \frac{2}{3} \tr F_W^4 \, ,
    \end{split}
\end{align}
where we recall that $k = 2n$ for type I ($\mathfrak{g}_W = \mathfrak{sp}(n)$), and $k=n$ in Sugimoto ($\mathfrak{g}_W = \mathfrak{so}(n)$) with a stack of $n$ 5-branes.

Summing these contributions, and using the behavior \eqref{eq:Pontryagin_decomposition} of Pontryagin classes under the tangent bundle decomposition ${\cal T}M|_{\Sigma} = {\cal T}\Sigma \oplus {\cal N}\Sigma$, we find
\begin{align}\label{eq:D5-matter-anomalies}
    \begin{split}
        I_8 & = \frac{k}{2} \left( \frac{3 p_1^2(R) - 4 p_2(R)}{192} + \frac{p_1(R) \, \tr F_B^2}{96} + \frac{\tr F_B^4}{24} \right) \\
        & \quad + \left( -\frac12 p_1(R) - \frac14 \tr F_B^2 + \frac{k}{2} \chi(N)\right) \left(\frac{k}{48} (2p_1(N) - p_1(R)) - \frac{\tr F_W^2}{2} \right) \, .
    \end{split}
\end{align}
Comparing with \eqref{eq:10d_factors_no-norm}, we see that the first line is $\frac{ck}{2} X_8 = n\,X_8$ (since $c=1$ for type I and $c=2$ for Sugimoto).

The origin of the second line is explained in the type I (and its heterotic dual) setting \cite{Dixon:1992if,Douglas:1995bn,Witten:1996hc,Mourad:1997uc} (see also \cite{Basile:2023knk}), essentially tracing back to a 6d Green--Schwarz term $\int_{\Sigma_6} B_2 \wedge Y_4$ on the 5-brane worldvolume where the 2-form is identified with the restriction of the 10d 2-form onto the worldvolume.
By properly taking into account the 5-brane source in the equation of motion for $H_3$ in spacetime, one can also explain the presence of the Euler class term, leading to the contribution $I_8^\text{(GS)} = - (X_4 + n \, \chi(N)) \,\wedge Y_4$ to the worldvolume anomaly.
Then the second line of \eqref{eq:D5-matter-anomalies} becomes
\begin{align}
\begin{split}
     & \left( -\frac12 p_1(R) - \frac14 \tr F_B^2 + \frac{k}{2} \chi(N)\right) \left(\frac{k}{48} (2p_1(N) - p_1(R)) - \frac{\tr F_W^2}{2} \right) \\
     \stackrel{\text{type I}}{=} \, & (X_4 + n \, \chi(N)) \Bigg( \underbrace{\frac{n}{12} p_1(N) - \frac{n}{24} p_1(R) - \frac{\tr F_W^2}{2}}_{=: Y_4^\text{(I)}} \Bigg) = - I_8^\text{(GS)} \, .
\end{split}
\end{align}
This gives the well-known inflow
\begin{align}
    I_8^\text{inflow} = - (I_8 + I_8^\text{(GS)} ) = - n \, X_8 \, .
\end{align}

For 5-branes in the Sugimoto model, the inflow works in the same fashion, except that we find slightly different numerical factors for the 4-form $Y_4$ in the 6d Green--Schwarz term:
\begin{align}\label{eq:Y4-5-brane-Sugimoto}
\begin{split}
     & \left( -\frac12 p_1(R) - \frac14 \tr F_B^2 + \frac{k}{2} \chi(N)\right) \left(\frac{k}{48} (2p_1(N) - p_1(R)) - \frac{\tr F_W^2}{2} \right) \\
     \stackrel{\text{Sugimoto}}{=} \, & \frac12 (X_4 + n \, \chi(N)) \Bigg( \frac{n}{24} p_1(N) - \frac{n}{48} p_1(R) - \frac{\tr F_W^2}{2} \Bigg) =: (X_4 + n \, \chi(N)) \wedge Y_4^\text{(S)} \\
     \Longrightarrow \quad & Y_4^\text{(S)} = \frac{n}{48} p_1(N) - \frac{n}{96} p_1(R) - \frac{\tr F_W^2}{4} \, .
\end{split}
\end{align}
It would be interesting to investigate a more ``microscopic'' derivation of this 4-form on the 5-branes in the Sugimoto model.

\section{Implications for the gauge group topology}
\label{sec:defect_spectra}

In this section we investigate how the spectra on the 1- and 5-branes inform the gauge group \emph{topology} in 10d spacetime.
In both type I and the Sugimoto model, the 10d massless fields are in the (anti-)symmetric tensor representations of the gauge algebra $\mathfrak{g}_B = \mathfrak{so}(32)$, $\mathfrak{sp}(16)$, which are invariant under the center of the respective simply-connected covers $Spin(32)$ and $Sp(16)$.
The basic question that we will answer is if the dynamical excitations localized on the branes changes this result.

As reviewed more recently \cite{Witten:2023snr}, it is known that stable non-BPS D0-branes in type I, which can be viewed as tachyon condensates from 1-brane/anti-1-brane pairs, carry the spinor representation of $Spin(32)$ \cite{Sen:1998tt, Sen:1998ki, Sen:1999mg}.
To see this, one has to quantize the massless fermions $\chi = \chi_{i=1,...,32}$ in the vector rep ${\bf 32}$ of $\mathfrak{g}_B = \mathfrak{so}(32)$ arising from open strings between the spacetime filling 9-branes and the 0-brane.
This gives $2^{16} = 2^{15} + 2^{15}$ states in the 0-brane Hilbert space, which decompose into the spinor and co-spinor (each of dimension $2^{15}$) of $\mathfrak{g}_B$.
Crucially, a single 0-brane has an $O(1) = \mathbb{Z}_2$ gauge symmetry which acts on the Hilbert space in the same way as one of the spinorial $\mathbb{Z}_2 \subset Z(Spin(32))$.
Therefore, the (worldline-)gauge invariant states of the 0-brane see the spacetime symmetry group $Spin(32)/\mathbb{Z}_2$ as a global symmetry.

In the following, we extend this chain of logic to study the gauge invariant spectrum of the 1- and 5-branes that are required by anomaly matching, as well as uncharged solitonic 4-branes in the Sugimoto model.
For type I, it confirms the above result, as expected from S-duality to the heterotic string.
For the Sugimoto model, where there are no stable 0-branes \cite{Sugimoto:1999tx}, the analysis serves as the first ``derivation'' (analogous to the type I ``derivation'') of the gauge group $Sp(16)/\mathbb{Z}_2$.

Note that this is not a full derivation (hence the quotation marks) as there are in principle additional solitonic 3-branes of the Sugimoto model \cite{Sugimoto:1999tx} which could host center-non-invariant states (thus fixing the gauge group to be $Sp(16)$).
Again, this is analogous to type I string theory, where other types of branes could potentially support states in the co-spinor representation of $\mathfrak{so}(32)$.
In the latter case, the gauge group is ultimately confirmed in the S-dual heterotic frame, where worldsheet arguments requires a self-dual character lattice which rules out $Spin(32)$ or $Spin(32)/(\mathbb{Z}_2 \times \mathbb{Z}_2)$.
In the Sugimoto model, we turn the logic around, and use the proposed $Sp(16)/\mathbb{Z}_2$ gauge group to conjecture a ``S-like'' duality to the web of non-supersymmetric heterotic string theories.

\subsection{\texorpdfstring{$\mathfrak{so}(32)$}{so(32)} representations on type I Green--Schwarz defects}

\subsubsection*{1-branes}

In type I, the relevant features of the 1-brane is very similar to those of the 0-brane.
Both have orthogonal Chan--Paton factors, which means in particular that the gauge symmetry on $n$ 1-branes includes the discrete factor $\mathbb{Z}_2 = O(1)$.
Moreover, for $n=1$ this $\mathbb{Z}_2$ acts on the open string modes on the 1-brane in the same fashion as the $O(1)$ on the 0-brane \cite{Polchinski:1995df}.
Therefore, the quantization of the 1-brane will also give rise to $\mathfrak{so}(32)$ spinor representations, one of which is projected out by the $\mathbb{Z}_2$ gauge symmetry.

Now, it is know that even numbers of 0-branes can decay, i.e., are not stable \cite{Sen:1998tt, Sen:1998ki, Sen:1999mg, Freed:1999vc}.
This is not true for 1-branes, but nevertheless we can show that they do not support $\mathfrak{so}(32)$-spinorial states in their Hilbert space.
Namely, for $2n =: N$ 1-branes the massless fermions we need to quantize is $\chi^{(3)} \equiv \chi_{i,\alpha}$ in \eqref{tab:D1 spectrum}, where $i = 1,...,32$ is the vector index of $\mathfrak{g}_B = \mathfrak{so}(32)_B$, and $\alpha = 1,...,2n=N$ the $\mathfrak{g}_W = \mathfrak{so}(N)_W$ vector index.
The quantization of such a Majorana--Weyl fermion amounts to impose the anti-commutation relation
\begin{align}
    \{ \chi_{i, \alpha}, \chi_{j, \beta} \} = \delta_{ij} \delta_{\alpha\beta} \, .
\end{align}
By rearranging the indices $(i,\alpha) \rightarrow I$, $I=1,..., 32N$, this can be brought into the canonical form
\begin{align}
    \{ \chi_I, \chi_J \} = \delta_{IJ}
\end{align}
for a Clifford algebra over the vector representation ${\bf 32 N}$ of $\mathfrak{so}(32N)$.
In fact, the existence of this rearrangement is tied to the special subalgebra
\begin{align}
    \mathfrak{so}(k \, \ell) \supset \mathfrak{so}(k) \oplus \mathfrak{so}(\ell) \, , \quad \boldsymbol{k \, \ell} \rightarrow (\boldsymbol{k}, \boldsymbol{\ell})
\end{align}
for any integer $k$, $\ell$.
Thus, the quantization of $\chi$ leads to $\mathfrak{so}(32N)$ spinors, which decomposes into $\mathfrak{so}(32)_B \oplus \mathfrak{so}(N)_W$ representations following the above branching.
We must then select the branched products with a singlet representation under $\mathfrak{so}(N)_W$ to find the gauge invariant sector.

The branching rules follow from the general surjective projection map $\pi: \Lambda_\text{r}^{\mathfrak{so}(k \ell)} \rightarrow \Lambda_\text{r}^{\mathfrak{so}(k)} \oplus \Lambda_\text{r}^{\mathfrak{so}(\ell)}$ between the root lattices associated with the above embedding.
By linearly extending it to the weight lattice $\Lambda_\text{w}^{\mathfrak{so}(k \ell)} \subset \Lambda_\text{r}^{\mathfrak{so}(k \ell)} \otimes \mathbb{Q}$, which is in fact generated by roots plus the spinorial highest weights $w_s$, $w_c$, we can then compute the branching rules for the other representations.
For our purposes of determining the (non-)invariance under the center of $Spin(32)_B$, it is easier to consider the induced projection
\begin{align}
\begin{split}
    \overline{\pi} : Z(Spin(32N)) = \, & \frac{\Lambda_\text{w}^{\mathfrak{so}(32N)}}{\Lambda_\text{w}^{\mathfrak{so}(32N)}} \longrightarrow \\
    \longrightarrow \quad & \frac{\Lambda_\text{w}^{\mathfrak{so}(32)} \oplus \Lambda_\text{w}^{\mathfrak{so}(N)}}{\Lambda_\text{r}^{\mathfrak{so}(32)} \oplus \Lambda_\text{r}^{\mathfrak{so}(N)}} = \frac{\Lambda_\text{w}^{\mathfrak{so}(32)}}{\Lambda_\text{r}^{\mathfrak{so}(32)}} \oplus \frac{\Lambda_\text{w}^{\mathfrak{so}(N)}}{\Lambda_\text{r}^{\mathfrak{so}(N)}} \\
    = \, & Z(Spin(32)) \times Z(Spin(N)) \, .
\end{split}
\end{align}
On the left-hand side, the generators for $Z(Spin(32N)) = \mathbb{Z}_2 \times \mathbb{Z}_2$ are the equivalence classes of $w_s$ and $w_c$.
Then, the possible irreps of $\mathfrak{so}(32)_B \oplus \mathfrak{so}(N)_W$ that come from the decomposition of the (co-)spinor are all in the same equivalence class of $\overline{\pi}([w_\text{s/c}]) \in Z(Spin(32) \times Spin(N))$.
The $\mathfrak{so}(N)_W$ invariant representations must then be those which are mapped to $0 \in Z(Spin(N))$.

A concrete representation of $\pi$ is displayed in \cite[eqs.~(6.100), (6.104) with $m=16$.]{Yamatsu:2015npn}.
The matrix there is presented in the Dynkin basis consisting of the fundamental weights $w_k$, $k =1, ..., \text{rank}(\mathfrak{g})$.\footnote{Recall that the fundamental weights $w_k$ are defined by a choice of simple roots $\alpha_{\ell}$ of the algebra $\mathfrak{g}$, such that $w_k \cdot \alpha^\vee_\ell = \delta_{k\ell}$, where ``$\cdot$'' denotes the inner product on the root space, and $\alpha_\ell^\vee = \frac{\alpha_\ell}{2 \alpha_\ell \cdot \alpha_\ell}$ is the $\ell$-th coroot.
We numerate the simple roots in the standard way associated with the Dynkin diagrams: for $\mathfrak{g} = \mathfrak{so}(2n)$ the two ``branched'' nodes of the D-type Dynkin diagram are $\alpha_{n-1}$ and $\alpha_n$, and for $\mathfrak{g} = \mathfrak{sp}(n)$ the \emph{long} root is $\alpha_n$.
}
As we only require the projection of the spinorial highest weights $w_s = w_{32N}$ and $w_c = w_{32N-1}$, these can be read of from the last two columns of the matrices in that reference, which leads to the following equations (which in fact holds for both even and odd $N$):
\begin{align}\label{eq:projection_so-to-so-weights}
    \begin{split}
        \pi(w_s) & = N \, w_s^{(32)} \, , \\
        \pi(w_c) & = w_c^{(32)} + (N-1) \, w_s^{(32)} + w_v^{(N)} \, ,
    \end{split}
\end{align}
where $w_{s/c}^{(32)}$ denote the spinorial highest weights of $\mathfrak{so}(32)$, and $w_v^{(N)} = w_1^{(N)}$ the vector highest weight of $\mathfrak{so}(N)$.
All of these are order two elements modulo roots, so for even $N$, we have
\begin{align}
    \begin{split}
        \overline\pi([w_s]) & = 2n [w_s^{(32)}] = (0,0) \in Z(Spin(32)) \times Z(Spin(N)) \, , \\
        \overline\pi([w_c]) & = [w_c^{(32)}] - [w_s^{(32)}] + 2n[w_s^{(32)}] + [w_v^{(N)}] = [w_v^{(32)}] + [w_v^{(N)}] \, ,
    \end{split}
\end{align}
where we used $w_s \pm w_c \equiv w_v \mod \Lambda_\text{r}^{\mathfrak{so}(2n)}$.
Since for even $N$, $[w_v^{(N)}] \neq 0 \in Z(Spin(N))$, we see that the requirement of $\mathfrak{so}(N)_W$ gauge invariance removes the second equivalence class.
This shows that for a stack of even 1-branes there are no $\mathfrak{so}(32)$ spinors, consistent with the claimed analogy to the 0-brane worldline theory.

For a sanity check, let us also discuss the odd $N$ case.
Here we have $[w_v^{(N)}] = 0$, but $N [w^{(32)}_{s/c}] = [w^{(32)}_{s/c}] \neq 0$, so \eqref{eq:projection_so-to-so-weights} leads to
\begin{align}\label{eq:proj_so-spinors_odd}
    \overline{\pi}([w_s]) = [w_s^{(32)}] \, , \quad \overline{\pi}([w_c]) = [w_c^{(32)}] \, .
\end{align}
This is the same result as on a single 0-brane, namely that the quantization of the bifundamental fermions leads to two spinor representations of $\mathfrak{so}(32)_B$.
In this case, we must again utilize the additional $\mathbb{Z}_2$ symmetry that is part of the worldsheet gauge symmetry, and which projects out one of the two spinors.

\subsubsection*{5-branes}

We now turn to stacks of $N$ 5-branes of type I, which have a worldvolume gauge symmetry $\mathfrak{g}_W = \mathfrak{sp}(N)_W$.
The fermion field that carry a non-trivial $\mathfrak{g}_B = \mathfrak{so}(32)_B$ representation is again the bifundamental $\varphi^{(3)} \equiv \varphi_{i, \alpha}$, $i=1,...,32$, $\alpha=1,...,2N$ in \eqref{tab:D5 spectrum}.
For the quantization of this fermion, it is important to recall that $\varphi$ is a sympletic-Majorana--Weyl spinor, since the naive anti-commutation relation
\begin{align}
    \{ \varphi_{i,\alpha}, \varphi_{j,\beta} \} = \delta_{ij} \omega_{\alpha\beta}
\end{align}
is not consistent due to the anti-symmetry of $\omega_{\alpha\beta} = -\omega_{\beta\alpha}$ which is the invariant symplectic tensor of $\mathfrak{sp}(N)_W$,
\begin{align}
    \omega_{\alpha\beta} = \begin{pmatrix}
        \underline{0} & \mathds{1}_{N} \\
        -\mathds{1}_{N} & \underline{0}
    \end{pmatrix}_{\alpha\beta} \, .
\end{align}

Instead, the symplectic-Majorana--Weyl condition effectively reduces the physical degrees of freedoms associated with the pseudo-real representation $({\bf 32,2N})$ to the real representation $({\bf 32,N})$ of $\mathfrak{so}(32)_B \oplus \mathfrak{so}(N)_W$.
It is tied to the special embedding $\mathfrak{sp}(N)_W \supset \mathfrak{sp}(1)_W \oplus \mathfrak{so}(N)_W$, ${\bf 2N} \rightarrow ({\bf 2}, {\bf N})$, where the $\mathfrak{sp}(1)_W \cong \mathfrak{su}(2)_W$ relates the two real ``halves'' such that the symplectic tensor of $\mathfrak{sp}(N)_W$ is given by $\omega = \epsilon \otimes \delta$, where $\epsilon$ and $\delta$ are the canonical invariant tensors of $\mathfrak{sp}(1)_W$ and $\mathfrak{so}(N)_W$, respectively.

Thus, we end up with having to quantize the fields $\varphi_{i,\alpha}$ in the $({\bf 32, N})$ of $\mathfrak{so}(32)_B \oplus \mathfrak{so}(N)_W$, which amounts to impose the now consistent anti-commutation relation
\begin{align}
    \{ \varphi_{i,\alpha}, \varphi_{j,\beta} \} = \delta_{ij} \delta_{\alpha\beta} \quad \text{with } \, i,j=1,...,32 \, \text{ and } \, \alpha,\beta =1,...,N \, . 
\end{align}
These components rearrange into the vector ${\bf 32N}$ of $\mathfrak{so}(32N)$, consistent with the existence of the special embedding
\begin{align}
    \mathfrak{so}(32N) \supset \mathfrak{so}(32)_B \oplus \mathfrak{so}(N)_W \, , \quad {\bf 32N} \rightarrow ({\bf 32},{\bf N}) \, .
\end{align}
This is now the same setting we quantized in the 1-brane case above, leading to the projection \eqref{eq:projection_so-to-so-weights}.

By the same logic, there are no non-trivial center-equivalence classes under $\mathfrak{so}(32)_B$ that survive in the $\mathfrak{so}(N)_W \subset \mathfrak{sp}(N)_W$-invariant sector for even $N$.
For odd $N$, the remaining $\mathfrak{sp}(1)_W$ factor now plays the role of the $O(1) \subset O(N)$ on the 1-branes, but in this case it projects out both spinor equivalence classes in \eqref{eq:proj_so-spinors_odd}.
To see this, consider, for simplicity, $N=1$.
Then the (zero modes of the) fermions $\varphi \equiv \varphi_{i=1,...,32}$ create the $\mathfrak{so}(32)_B$ spinors when acting on the Hilbert space vacuum.
As symplectic Majorana--Weyl fermions, an $\mathfrak{sp}(1)_W$ doublet is formed out of two copies of $\varphi_i$.
Then an $\mathfrak{sp}(1)_W$ singlet is formed by contracting two doublets using the symplectic form (i.e., epsilon tensor), leading to $\varphi_i \varphi_j - \varphi_j \varphi_i = [\varphi_i, \varphi_j] \equiv S_{ij}$, which is a generator of the Lie algebra $\mathfrak{so}(32)_B$, i.e., in the adjoint representation of $\mathfrak{g}_B$.

\subsection{\texorpdfstring{$\mathfrak{sp}(16)$}{sp(16)} representations on Green--Schwarz defects in Sugimoto model}

Let us now turn to the Sugimoto model.
The procedure is almost identical, requiring replacing orthogonal with symplectic algebras.
Because this swaps the role of gauge and flavor symmetries on the 5-brane, the analysis there will be slightly different than in the type I case.

\subsubsection*{1-branes}

The relevant field $\chi_3 \equiv \chi$ iHeten \eqref{tab:D1 spectrum} of on the 1-brane of the Sugimoto model is in the bifundamental representation $({\bf 32}, {\bf 2n})$ of $\mathfrak{sp}(16)_B \oplus \mathfrak{sp}(n)_W$.
The quantization of these Weyl fermion $\chi \equiv \chi_{i, \alpha}$ amounts to impose the anti-commutation relation
\begin{align}
    \{ \chi_{i,\alpha}, \chi_{j,\beta} \} = \omega_{ij} \omega_{\alpha\beta} \, ,
\end{align}
where $i=1,...,32$ is the index for the fundamental of $\mathfrak{sp}(16)_B$, and $\alpha=1,...,2n$ that of $\mathfrak{sp}(n)_W$, and $\omega$ is the invariant symplectic tensor of the respective algebras,
\begin{align}
    \omega_{ij} = \begin{pmatrix}
        \underline{0} & \mathds{1}_{16} \\
        -\mathds{1}_{16} & \underline{0}
    \end{pmatrix}_{ij} \, , \quad
    \omega_{\alpha\beta} = \begin{pmatrix}
        \underline{0} & \mathds{1}_n \\
        - \mathds{1}_n & \underline{0}
    \end{pmatrix}_{\alpha\beta} \, .
\end{align}
By suitable rearranging the $32 \times 2n$ possible values for $(i,\alpha)$ into a new index $I=1,..., 64n$, the anti-commutation relation can be brought into the canonical form
\begin{align}
    \{ \chi_I, \chi_J \} = \delta_{IJ}
\end{align}
for a Clifford algebra over the vector representation ${\bf 64n}$ of $\mathfrak{so}(64n)$.
This rearranging is associated with the special branching
\begin{align}
    \mathfrak{so}(4 m n) \supset \mathfrak{sp}(m) \oplus \mathfrak{sp}(n) \, , \qquad {\bf 4 m n} \rightarrow ({\bf 2m}, {\bf 2 n}) \, ,
\end{align}
of $\mathfrak{so}(4mn)$.
Thus the Hilbert space of the 1-branes carry the spinor representations of $\mathfrak{so}(64n)$, whose decomposition along the above branching gives the possible $\mathfrak{sp}(16)_B \oplus \mathfrak{sp}(n)_W$ irreducible representations.

To compute these branchings, we use the concrete projection matrix $\pi$ displayed in \cite[eq.~(6.106)]{Yamatsu:2015npn} with $m=16$, whose action on the spinorial fundamental weights, $w_s = w_{64n}$ and $w_c = w_{64n-1}$ are again in just last two columns of the matrix:
\begin{align}
    \begin{split}
        \pi(w_c) & = w_{15}^{\mathfrak{sp}(16)} + (n-1) \, w_{16}^{\mathfrak{sp}(16)} + w_1^{\mathfrak{sp(n)}} \, , \\
        \pi(w_s) & = n \, w_\text{16}^{\mathfrak{sp}(16)} \, .
    \end{split}
\end{align}

To understand this result, we recall the following basic facts about the weight system of $\mathfrak{sp}(n)$.
First, given its Cartan matrix $C_{ij} = \alpha_i \cdot \alpha_j^\vee$, the inverse Cartan matrix
\begin{align}
    (C^{-1})_{ij} = \begin{cases}
        \text{min}(i,j) \, , & j < n \, ,\\
        \frac{i}{2} \, , & j = n \, ,
    \end{cases}
\end{align}
encodes in its rows the fundamental weights,
\begin{align}
    w_k = (C^{-1})_{k\ell} \, \alpha_\ell \quad \Leftrightarrow \quad w_k \cdot \alpha_\ell^\vee = \delta_{k\ell} \, .
\end{align}
From this, we see that
\begin{align}
    \begin{split}
        \left[ w_{15}^{\mathfrak{sp}(16)} \mod \Lambda_\text{r}^{\mathfrak{sp}(16)} \right] &= 1 \in \mathbb{Z}_2 = \Lambda_\text{w}^{\mathfrak{sp}(16)} / \Lambda_\text{r}^{\mathfrak{sp}(16)} \, , \\
        \left[ w_{16}^{\mathfrak{sp}(16)} \mod \Lambda_\text{r}^{\mathfrak{sp}(16)} \right] &= 0 \in \mathbb{Z}_2 = \Lambda_\text{w}^{\mathfrak{sp}(16)} / \Lambda_\text{r}^{\mathfrak{sp}(16)} \, , \\
        \left[ w_{1}^{\mathfrak{sp}(n)} \mod \Lambda_\text{r}^{\mathfrak{sp}(n)} \right] &= 1 \in \mathbb{Z}_2 = \Lambda_\text{w}^{\mathfrak{sp}(n)} / \Lambda_\text{r}^{\mathfrak{sp}(n)} \, , \\
        \Rightarrow \quad \overline{\pi}([w_c]) & = (1,1) \in Z(Sp(16)) \times Z(Sp(n)) \, , \\
        \overline{\pi}([w_s]) & = (0,0) \in Z(Sp(16)) \times Z(Sp(n)) \, .
    \end{split}
\end{align}

This means that co-spinor of $\mathfrak{so}(64n)$ cannot project onto any $\mathfrak{g}_W$ singlets.
Instead, any gauge invariant state can only arise in the decomposition of the spinor class.
While this analysis here does not show that such irreps actually do arise, the relevant consequence is that any such irrep \emph{must} also be invariant under the center of the spacetime symmetry group $Sp(16)$.
This shows that the quantization of the effective gauge theory on the 1-brane stack can only produce gauge invariant excitations that transform under the spacetime gauge symmetry group $G_B = Sp(16)/\mathbb{Z}_2$.

\subsubsection*{5-branes}

We now turn to the spectrum on a stack of $n$ 5-branes, which as the 1-branes in type I have, in addition to the continuous gauge symmetry with algebra $\mathfrak{so}(n)_W$, also a discrete $\mathbb{Z}_2$ factor.
Just as in the type I case, we must account for the symplectic-Majorana--Weyl nature of the relevant field $\varphi_3 \equiv \varphi$ which is in the bifundamental representation of $\mathfrak{sp}(16)_B \oplus \mathfrak{so}(n)_W$.
However, the difference in the Sugimoto model is that the $\mathfrak{sp}(1)$ factor that doubles up the degrees of freedom comes from the bulk symmetry $\mathfrak{g}_B = \mathfrak{sp}(16)_B \supset \mathfrak{sp}(1)_B \oplus \mathfrak{so}(16)_B$, which from the worldvolume perspective is a global rather than a gauge symmetry.

Keeping this in mind, we first proceed with the quantization of $\varphi_{i,\alpha}$, with $i=1,...,16$ the $\mathfrak{so}(16)_B$ vector index, and $\alpha = 1,..., n$ the $\mathfrak{so}(n)_W$ vector index, by imposing
\begin{align}
    \{ \varphi_{i,\alpha}, \varphi_{j,\beta} \} = \delta_{ij} \delta_{\alpha\beta} \, .
\end{align}
Rearranging again into $\mathfrak{so}(16n)$ vector indices, we find the associated Clifford algebra, whose representation space (which is a subspace of the Hilbert space carrying any $\mathfrak{g}_B$-non-invariant states) is generated by the spinor and co-spinor representations.
These can be decomposed with respect to the special embedding $\mathfrak{so}(16n) \supset \mathfrak{so}(16)_B \oplus \mathfrak{so}(n)_W$ analogously to \eqref{eq:projection_so-to-so-weights}, i.e.,
\begin{align}\label{eq:proj_sugimoto_5-brane}
    \overline\pi ([w_s]) = \left\{ \begin{array}{l l}
        0 \, , & n \, \text{ even,} \\ \relax
        [w_s^{(16)}] \, , & n \, \text{ odd} 
    \end{array} \right\} \, , \qquad \overline\pi ([w_c]) = \left\{ \begin{array}{l l}
        [w_v^{(16)}] + [w_v^{(n)}] \, , & n \, \text{ even,} \\ \relax
        [w_c^{(16)}] \, , & n \, \text{ odd} 
    \end{array} \right\} \, .
\end{align}

Imposing $\mathfrak{so}(n)_W$-invariance here gives certain $\mathfrak{so}(16)_B$ (center-equivalence class of) representations $\{\rho_x\}$.
To reassemble these into an $\mathfrak{sp}(16)_B$-irrep ${\bf R}$ requires to build the suitable $\mathfrak{so}(16)_B$-irreps ${\bf r}_y$ out of the $\rho_x$ such that ${\bf R} \rightarrow \bigoplus_y ({\bf s}_y, {\bf r}_y)$ under $\mathfrak{sp}(16)_B \supset \mathfrak{sp}(1)_B \oplus \mathfrak{so}(16)_B$.
Again we find a simplification if we focus on the center-equivalence classes that can arise from the projection
\begin{align}\label{eq:projection_center-equiv-class_sp16}
\begin{split}
    \overline{\pi}_B: \mathbb{Z}_2 = \frac{\Lambda_\text{w}^{\mathfrak{sp}(16)}}{\Lambda_\text{r}^{\mathfrak{sp}(16)}} & \rightarrow \frac{\Lambda_\text{w}^{\mathfrak{sp}(1)}}{\Lambda_\text{r}^{\mathfrak{sp}(1)}} \oplus \frac{\Lambda_\text{w}^{\mathfrak{so}(16)}}{\Lambda_\text{r}^{\mathfrak{so}(16)}} = \mathbb{Z}_2 \times (\mathbb{Z}_2 \times \mathbb{Z}_2) \, , \\
    1 = [{\bf 32}] & \mapsto ( [{\bf 2}], [{\bf 16}] ) = (1, (1,1)) \, .
\end{split}
\end{align}
This means that in the branching ${\bf R} \rightarrow \bigoplus_y ({\bf s}_y, {\bf r}_y)$, the $\mathfrak{so}(16)_B$ irreps ${\bf r}_y$ must be either in the trivial equivalence class (if ${\bf R}$ is in the trivial class), or in the vector equivalence class (if ${\bf R}$ is in the non-trivial class).

Putting things together, we see first that for even $n$, the $\mathfrak{so}(n)_W$-invariance condition restricts to states which are in the trivial equivalence class under $\mathfrak{so}(16)_B$, which in turn means that the associated $\mathfrak{sp}(16)_B$ representations that can be constructed from these also transform trivially under $Z(Sp(16))$.

For odd $n$, we must further make use of the discrete $O(1) = \mathbb{Z}_2$ symmetry that is part of the gauge symmetries on the 5-brane stack.
Just as on the 0- and 1-branes in type I, its odd action on the fermion fields $\varphi_{i,\alpha}$ in the $({\bf 16, n})$ of $\mathfrak{so}(16)_B \oplus \mathfrak{so}(n)_W$ means that the $\mathbb{Z}_2$ invariant subspace of the Hilbert space only contains (tensor powers of) one of two spinor representations of $\mathfrak{so}(16)_B$.
Since the vector-equivalence class of $\mathfrak{so}(16)_B$ can only be obtained by tensoring the two different spinors with each other, projecting out one of them means that the Hilbert space does not contain suitable $\mathfrak{so}(16)_B$ representations to construct $\mathfrak{sp}(16)_B$ representations in a non-trivial center-equivalence class, due to \eqref{eq:projection_center-equiv-class_sp16}.
Therefore, we find that the 5-brane-stack's gauge invariant states transform in representations of $Sp(16)/\mathbb{Z}_2$.

\subsection{A brief look at 4-branes in the Sugimoto model}

The analysis of odd numbers of 5-branes carry over also to the stable 4-branes of the Sugimoto model, which in terms of K-theory charges \cite{Witten:1998cd} are the analogues of 0-branes in type I, by virtue of Bott periodicity \cite{Sugimoto:1999tx, Dudas:2001wd}.
Thus, while even numbers of 4-branes decay, a single 4-brane can give rise to stable 10d excitations.
The representations under $\mathfrak{sp}(16)_B$ follow the same construction as with 5-branes, because the worldvolume gauge symmetry is $O(1) = \mathbb{Z}_2$, and the massless open string modes are bifundamental symplectic Majorana fermions.
Their quantization is analogous to the 5-brane setting, and leads to only $\mathfrak{sp}(16)_B$-center-invariant representations in the gauge invariant sector, further lending evidence to the 10d gauge group $Sp(16)/\mathbb{Z}_2$.

\subsection{Evidence for duality to non-SUSY heterotic strings}
\label{sec:duality_evidence}

Having the gauge group topology $Sp(16)/\mathbb{Z}_2$ opens up another interesting feature for the Sugimoto model, namely, to be a \emph{potential} dual to (one of) the non-supersymmetric heterotic string theories (see also \cite{Basile:2022zee}).
Again, we draw inspiration from their supersymmetric cousins, where the type I string is famously S-dual to the supersymmetric heterotic $Spin(32)/\mathbb{Z}_2$ theory.

While there are various physical aspects that corroborate this duality \cite{Polchinski:1995df}, one concerns the gauge group topology $G$ of the $\mathfrak{g} = \mathfrak{so}(32)$ gauge symmetry which is present on both sides of the duality.
As already mentioned above, the heterotic worldsheet description of this theory requires a self-dual charge lattice.
In more careful terms, this means that the \emph{character} lattice $\Lambda_c^G$ of the $G$ gauge theory is the same as its \emph{cocharacter} lattice $\Lambda_\text{cc}^G$.\footnote{We are adopting the nomenclature where, for a simple Lie algebra $\mathfrak{g}$, the weight and coweight lattices, $\Lambda^\mathfrak{g}_\text{w}$ and $\Lambda^\mathfrak{g}_\text{cw}$, are duals of the coroot and root lattices, $\Lambda^\mathfrak{g}_\text{cr}$ and $\Lambda^\mathfrak{g}_\text{r}$, respectively.
An associated Lie group $G$ is specified by the (co-)character lattice $\Lambda^G_c \subset \Lambda^\mathfrak{g}_\text{w}$ ($\Lambda^G_\text{cc} \subset \Lambda^\mathfrak{g}_\text{cw}$, with $(\Lambda^G_\text{cc})^* = \Lambda^G_c$) with $Z(G) = \Lambda_c^G / \Lambda_\text{r}^\mathfrak{g} = \Lambda_\text{cw}^\mathfrak{g} / \Lambda_\text{cc}^G$ and $\pi_1(G) = \Lambda_\text{cc}^G/\Lambda_\text{cr}^\mathfrak{g} = \Lambda_\text{w}^\mathfrak{g} / \Lambda_c^G$; if $G$ is simply-connected, then $\Lambda^G_c = \Lambda^\mathfrak{g}_\text{w}$ and $\Lambda_\text{cc}^G = \Lambda_\text{cr}^\mathfrak{g}$.
In the literature $\Lambda^G_c$ ($\Lambda_\text{cc}^G$) is also oftentimes called the ``(co-)weight lattice of $G$''.
}
As the usual S-duality transformation exchanges root and character lattices with coroot and cocharacter lattices, respectively, we see that the group structure $Spin(32)/\mathbb{Z}_2$ seen by the D0-branes of type I is indeed the only possibility compatible with the S-dual heterotic theory.

Applying the same logic to the Sugimoto model, we first identify the ``S-dual'' gauge algebra as $\mathfrak{so}(33)$, since
\begin{align}
    \begin{split}
        \Lambda_\text{r}^{\mathfrak{sp}(16)}= \Lambda_\text{cr}^{\mathfrak{so}(33)} \quad \text{and} \quad \Lambda_\text{cr}^{\mathfrak{sp}(16)} = \Lambda_\text{r}^{\mathfrak{so}(33)} \, .
    \end{split}
\end{align}
Then, the putative gauge group $Sp(16)/\mathbb{Z}_2$ would be dualized to $\hat{G} = Spin(33)$ because
\begin{align}
\begin{split}
    \Lambda_\text{cc}^{\hat{G}} &\stackrel{!}{=} \Lambda_c^{Sp(16)/\mathbb{Z}_2} = \Lambda_\text{r}^{\mathfrak{sp}(16)} = \Lambda_\text{cr}^{\mathfrak{so}(33)} \, , \\
    \Lambda_c^{\hat{G}} &\stackrel{!}{=} \Lambda_\text{cc}^{Sp(16)/\mathbb{Z}_2} = \Lambda_\text{cw}^{\mathfrak{sp}(16)} = (\Lambda_\text{r}^{\mathfrak{sp}(16)})^* = (\Lambda_\text{cr}^{\mathfrak{so}(33)})^* = \Lambda_\text{w}^{\mathfrak{so}(33)} \, .
\end{split}
\end{align}

While this algebra/group does not appear as the physical gauge symmetry of either of the non-supersymmetric heterotic theories \cite{Alvarez-Gaume:1986ghj, Dixon:1986iz, Seiberg:1986by, Kawai:1986vd}, these $\mathfrak{so}(33)$ lattices encode the charges of the $Spin(32)$ heterotic string \cite{Fraiman:2023cpa}.
Namely, via the embedding $\mathfrak{so}(33) \supset \mathfrak{so}(32)$ where the adjoint decomposes as ${\bf 528} \rightarrow {\bf 496} \oplus {\bf 32}$, the long $\mathfrak{so}(33)$ roots gives rise to the gauge bosons of $\mathfrak{so}(32)$, the \emph{short} roots become the spacetime tachyons in the vector representation.
The additional vectors in $\Lambda_\text{w}^{\mathfrak{so}(33)} = \Lambda_c^{Spin(33)}$ are in the equivalence class of the spinor representation, which decompose into the sum of the two spinors of $\mathfrak{so}(32)$; these are massive states in the $Spin(32)$ heterotic theory.
The fact these states appear in the cocharacter rather than the character lattice of the Sugimoto model suggests that there may indeed be a non-supersymmetric version of the duality between type I and heterotic.

Clearly, a simple agreement of the lattices is far from a conclusive proof for the existence of such a duality.
One obvious question that needs to be addressed is how the distinction between gauge bosons and tachyons in the heterotic model arises in this duality.
A related, more conceptual question is how the different types of vacuum instability --- the tachyons in the heterotic model and the running dilaton potential in the Sugimoto model --- are mapped onto each other.
A fruitful direction could be to explore the duality in lower dimensions, where one can obtain (at least perturbatively) stable Minkowski vacua \cite{Dudas:2000ff, Basile:2018irz}.
In fact, the $\mathfrak{so}(33)$ lattices appear more directly as an infinite distance limit in the moduli space which connects all non-supersymmetric heterotic strings on an $S^1$ \cite{Fraiman:2023cpa}, suggesting a careful analysis of circle reductions of the Sugimoto model to shed more light the duality, and perhaps establish a connection to earlier works on dualities between non-supersymmetric 10d strings \cite{Blum:1997cs, Blum:1997gw}.

\section{Bottom-up constraints for \texorpdfstring{$Sp(16)/\mathbb{Z}_2$}{Sp(16)/Z2} gauge theories}
\label{sec:bottom-up}

In this section we explore anomaly inflow as a bottom-up constraint on 10d $Sp(16)/\mathbb{Z}_2$ gauge theories coupled to gravity.
In the absence of supersymmetry, there is a priori no restriction on the massless matter content of the theory aside from having only center-invariant representations and the cancellation of all gauge anomalies with a Green--Schwarz mechanism involving a 2-form gauge field.
By the Completeness Hypothesis \cite{Polchinski:2003bq, Banks:2010zn}, there must then exist dynamical 1- and 5-branes charged under this gauge field, whose internal dynamics must cancel the anomaly inflow.
A putatively consistent spacetime chiral spectrum may therefore be ruled out if no suitable defect worldvolume dynamics exists to match the anomaly inflow.

From a bottom-up perspective, the inflow specifies a characteristic polynomial $I_n$, $n=4, 8$ for the 1- and 5-brane, respectively.
The consistency condition is then that $I_n$ can be expressed as the anomaly polynomial of a sum virtual irreducible representations under the spacetime gauge symmetry group (and the Poincar\'e symmetry along the defect).
The physical intuition is that the localized chiral degrees of freedom transform in this representation under the spacetime gauge symmetry, which is a global symmetry on the defect.
This applies also to cases where the defect worldvolume theory has no weakly-coupled Lagrangian description, as exemplified in the case of NS5-branes in the $SO(16)^2$ non-supersymmetric heterotic string \cite{Basile:2023knk}.

For this to be a practically feasible consistency condition, we must make certain simplifying assumptions.
First, due to computational limitations we can only consider a finite set of different representations under the spacetime gauge symmetry.
We then parametrize the model via the net chiral numbers of spacetime fermions in these representations as well as the chiral index of the virtual fermion representations on the 1- and 5-branes, such that the inflow from the spacetime representations matches that of the virtual representations on the branes.
Furthermore, we allow for a contribution to $I_8$ on the 5-branes coming from a 6d Green--Schwarz mechanism; we assume that this contribution is the same as in the Sugimoto model \eqref{eq:Y4-5-brane-Sugimoto}, and leave the study with more general possibilities for future works.
Note that this analysis does not depend on the worldvolume gauge symmetry, and only on the net number of virtual chiral contributions in different spacetime gauge representations.

To begin with, we start with a general 10d spectrum containing singlets and gravitinos as well as 2-index representations which are all invariant under the center of $Sp(16)$.
We parametrize these their net chirality as
\begin{equation}
    \begin{split}
        n_\textbf{1} \quad & \text{singlets},\\
        n_g \quad & \text{gravitinos},\\
        n_{\text{\tiny{\yng(1,1)}}} \quad & \text{antisymmetric representations},\\
        n_{\text{\tiny{\yng(2)}}}\quad & \text{symmetric representations}.
    \end{split}
    \label{eq:Number convention}
\end{equation}
The spacetime anomaly polynomial is then
\begin{equation}
    I_{12} = n_\textbf{1} I_{12\,\textbf{1}} + n_g I_{12\,g} + n_{\text{\tiny{\yng(1,1)}}} I_{12}{}_{\text{\tiny{\yng(1,1)}}} + n_{\text{\tiny{\yng(2)}}} I_{12\,\text{\tiny{\yng(2)}}} \, .
    \label{eq:General rank 2 anomaly polynomial}
\end{equation}

The consistency of these models requires the Green--Schwarz mechanism to cancel the anomaly, i.e., $I_{12}$ must factorize.
This is obstructed by a non-zero $\tr F^6$ term, which only receives contributions from the symmetric representation.
Therefore,
\begin{equation}
    n_{\text{\tiny{\yng(2)}}} = 0.
\end{equation}
Furthermore, the coefficient of $\tr R^6$ must cancel, leading to
\begin{equation}
    n_\textbf{1} = 495 n_g - 496 n_{\text{\text{\tiny{\yng(1,1)}}}}.
    \label{eq:10d rank 2 number of singlets}
\end{equation}
At this point we must require that (\ref{eq:General rank 2 anomaly polynomial}) factorize in the form
\begin{equation}
    I_{12} = \tilde{X}_4 \wedge \tilde{X}_8\,,
    \label{eq:Bulk anomaly polynomial factorization}
\end{equation}
where the tilde denotes that these factors will generically differ from their version in the Sugimoto model.
Indeed, $\tilde{X}_4$ is a general $4$-form of the form
\begin{equation}\label{eq:X4_bottom-up}
    \tilde{X}_4 = \frac{1}{2}\left[\tr R^2 - \beta \tr F^2\right] \, .
\end{equation}
Terms like $(\tr{F})^2$ do not appear since odd powers of the field strength vanish for $\mathfrak{sp}(n)$ algebras. 

We find that the anomaly polynomial factorizes in three different cases. The first solution, with $n_g = 0 $ and $ n_{\text{\tiny{\yng(1,1)}}} = 0$, has no chiral field content. 
The second case forces $n_{\text{\tiny{\yng(1,1)}}} = 0$ and $\beta = 0$, implying the absence of gauge-charged objects. 
In the third case which is physically more interesting, spacetime anomaly cancellation occurs when
\begin{equation}
    n_g = n_{\text{\tiny{\yng(1,1)}}} \quad \text{and} \quad \beta = 1 \, .
    \label{eq:10 rank 2 factorization condition}
\end{equation}
This fixes the anomaly inflow onto the 1-brane to be the same as the Sugimoto model, see \eqref{eq:10d_factors_no-norm}.

Thus, from an effective field theory perspective, any chiral number of anti-symmetric fields seem to give a consistent 10d theory, since the resulting anomaly can be cancelled by a Green--Schwarz mechanism.
However, as we shall see now, consistent inflow onto the charged defects can serve as a method to distinguish their consistency as a viable effective theory of quantum gravity.

\subsection{Constraints from brane anomaly}

On the branes, we parametrize the chiral degrees of freedom in terms of their net chiral number $n_{\cal R}$ in the virtual representation ${\cal R}$ that can be singlets, fundamentals, and 2-index (anti-)symmetric representations of $\mathfrak{sp}(16)$.
For simplicity, we just look at a stack of $n=1$ branes.

For the 1-brane, where the anomaly polynomial $I_4$ only receives contributions from these virtual representations, imposing a consistent inflow matching $I_4 = -\tilde{X}_4$ leads to 
\begin{equation}
    \begin{aligned}
        32 n_{\text{\tiny\yng(1)}} + n_{\textbf{1}}  + 496 n_{\text{\tiny\yng(1,1)}} +& 528 n_{\text{\tiny\yng(2)}}  = 24\,,\\
        n_{\text{\tiny\yng(1)}} + 30 n_{\text{\tiny\yng(1,1)}} + 34 n_{\text{\tiny\yng(2)}} &= 1\,,
    \end{aligned}   
    \label{eq:System_USp(32)_NOG_W}
\end{equation}
which admit as solution
\begin{equation}
    \begin{aligned}
    n_{\text{\tiny\yng(1)}} &= 1 + 2c_1 \,,\\ 
    n_{\textbf{1}} &= 504 + 208 c_1 + 512 c_2\,,\\ 
    n_{\text{\tiny\yng(1,1)}} &= -17-8c_1 - 17 c_2\,,\\ 
    n_{\text{\tiny\yng(2)}} &= 15 + 7 c_1 +15 c_2\,.
    \end{aligned}
\end{equation}
The choice $c_1 = 0$ and $c_2 = -1$ corresponds to the spectrum of the Sugimoto model.

In order to extract further constraints on the set of theories under consideration we require anomaly cancellation via inflow on the worldvolume of the 5-brane. The factorization of the bulk anomaly polynomial (\ref{eq:Bulk anomaly polynomial factorization}) gives us the expression for $\tilde{X}_8$, that, together with the conditions (\ref{eq:10 rank 2 factorization condition}), yields:
\begin{equation}
\tilde{X}_8 = \frac{1}{384}n^B_{\text{\tiny{\yng(1,1)}}}\left[\frac{1}{4}(\tr{R^2})^2 + \tr{R^4} - \tr{R^2}\tr{F^2} + 8 \tr{F^4}\right],
    \label{eq:Tilde X8}
\end{equation}
where $n^B_{\text{\tiny{\yng(1,1)}}}$ denotes the chiral number of antisymmetric representations of $Sp(16)/\mathbb{Z}_2$ in the 10d bulk. The anomaly inflow with one 5-brane is of the form
\begin{equation}
    -I_8 = \tilde{X}_8 + \tilde{X}_4\wedge Y_4.
\end{equation}
Again, we impose the anomaly inflow to cancel the brane anomaly. This happens when
\begin{equation}
    n_{\text{\tiny\yng(1)}} = 1 \, , \quad n_{\textbf{1}} = -2 \, , \quad n_{\text{\tiny\yng(1,1)}} = 0 \, , \quad n_{\text{\tiny\yng(2)}} = 0 \, , \quad n^B_{\text{\tiny\yng(1,1)}} =1 \, .
\end{equation}
This result is interesting from both bulk and brane perspective. The spectrum on the brane is consistent with the 5-brane spectrum in the Sugimoto.
Indeed, as we can read off from (\ref{tab:D5 spectrum}), the net number of spacetime singlets degrees of freedom with $n=1$ is exactly $-2$ since
\begin{equation}
    2\left(\frac{n(n-1)}{2} - \frac{n(n+1)}{2} \right) = -2.
\end{equation}
Crucially, the brane analysis forces the net number of antisymmetric representations in spacetime to be one. This implies a spacetime spectrum of the form
\begin{equation}
    n_\textbf{1} = -1\,,\quad n_g = 1 \,, \quad n_{\text{\tiny{\yng(1,1)}}} = 1\,, \quad n_{\text{\tiny{\yng(2)}}} = 0 \, ,
\end{equation}
which is the same chiral spectrum as the Sugimoto model.

In summary, we find that a consistent $Sp(16)/\mathbb{Z}_2$ gauge theory with massless chiral fermions in up to rank-2 representations must have the same chiral spectrum as the Sugimoto model if the charged defects participating in the Green--Schwarz mechanism also host only up to rank-2 $\mathfrak{sp}(16)$ representations.

\subsection{Consistency of theories with higher rank representations}
\label{sec:Higher rank particles}
Following the same logic, we want to push forward our search for tentative $Sp(16)/\mathbb{Z}_2$ gauge theories by further enlarging the set of possible representations in both spacetime and brane woldvolume.
As odd rank tensor representations in spacetime are not invariant under the center $\mathbb{Z}_2$, we restrict our focus to certain even-rank representations.
Again, one could in principle allow their presence on the worldvolume, thus exploring further possibilities, but for simplicity we consider the identical sets for both spacetime and brane worlvolume.
The coefficients related to the traces in these representations that are crucial for anomaly polynomial computations can be found in appendix \ref{sec:Appendix higher rank trace coefficient}.

Our analysis starts with adding the rank-4 completely-symmetric representation
\begin{equation*}
    \yng(4)
\end{equation*}
to the previous spacetime spectrum. 
By imposing the factorization of $I_{12}$ and anomaly cancellation via Green--Schwarz mechanism, we find that the presence of such chiral degrees of freedom is an obstruction to the required factorization of the anomaly polynomial. This implies that there are no consistent $Sp(16)/\mathbb{Z}_2$ theories with chiral fermions in the rank-4 symmetric representation in addition to fundamentals and 2-index representations.

The second possibility we want to consider is to add the rank-4 completely antisymmetric representation
\begin{equation*}
    \yng(1,1,1,1)
\end{equation*}
together with the spectrum \eqref{eq:Number convention}. 
We find that the only non-trivial configuration to allow for a spacetime Green--Schwarz mechanism requires
\begin{equation}
    \beta = \frac{4095}{3071} 
\end{equation}
in \eqref{eq:X4_bottom-up}.
However, this result is forbidden by the anomaly cancellation via inflow on the 1-brane. Indeed, even allowing all the higher rank representations considered in this section on the worldvolume, there is no way to match the brane anomaly polynomial with the anomaly inflow given by the spacetime solution. 
Once again, we conclude that $Sp(16)/\mathbb{Z}_2$ theories featuring chiral rank-4 antisymmetric fermions are forbidden by anomaly inflow arguments.

We can also consider the rank-6 completely symmetric representation,
\begin{equation*}
    \yng(6) \, ,
\end{equation*}
together with the representations in (\ref{eq:Number convention}).
This has no spacetime anomaly for
\begin{equation}
    \beta = -2 \quad \text{and} \quad \beta = -4\,.
\end{equation}
Both options would in principle allow a consistent cancellation via inflow on 1-branes, but are forbidden by consistency on the 5-branes, at least with the 6d Green--Schwarz term \eqref{eq:Y4-5-brane-Sugimoto} from the Sugimoto model.

\section{Conclusions}
\label{sec:conclusions}

In this work, we focused on the role of 1- and 5-branes as a crucial bridge between the consistency of anomaly cancellation and the global gauge group structure in the Sugimoto model. 
Starting from the Green--Schwarz mechanism, we demonstrated how the $Sp(16)/\mathbb{Z}_2$ gauge group naturally emerges.
This insight not only solidifies the anomaly cancellation mechanism but also opens avenues for exploring potential dualities with non-supersymmetric heterotic string theories. 
Such dualities, if rigorously established, could provide deeper insights into the landscape of consistent non-supersymmetric string models and their low-energy effective descriptions. 
Furthermore, the bottom-up approach applied here demonstrates the utility of anomaly inflow matching as a viable tool to constrain effective gauge theories coupled to gravity without supersymmetry, though certain restrictive assumptions about the worldvolume of the defects must be made.

Looking forward, several directions merit further exploration. 
One interesting question concerns the worldvolume theory of the 5-branes in the Sugimoto model.
Drawing inspiration from the supersymmetric type I cousins, which are known to give rise to 6d little string theories \cite{Aspinwall:1996vc, Intriligator:1997kq, Blum:1997fw, Blum:1997mm}, it is intriguing to speculate whether the 5-branes of the Sugimoto model can provide an explicit realization to non-supersymmetric versions of such exotic UV-complete 6d theories.
If so, then the worldvolume anomaly found in Section \ref{sec:anomaly_inflow} along with the 4-form $Y^{(S)}_4$ necessary for the 6d GS-mechanism (which differs from type I) may play a vital role in their understanding.

Another exciting direction is to scrutinize possible dualities to non-supersymmetric heterotic strings, which parallel the well-known supersymmetric S-duality between heterotic strings and type I.
We have provided evidence for such dualities, based on our results on the global form of the gauge group in the Sugimoto model and analogous results (though based on different methods) in non-supersymmetric heterotic models \cite{Fraiman:2023cpa}.
As highlighted in Section \ref{sec:duality_evidence}, this would-be duality needs to account for the tachyonic states present on the heterotic side which, based on the supersymmetric intuitions, are expected to be mapped onto ``non-perturbative'' instabilities in the Sugimoto model.
A crucial step in further corroborating this duality is therefore understanding the detailed relationship between the heterotic tachyons and the running dilaton potential of the Sugimoto model.

Finally, the ``uniqueness'' of the Sugimoto model as the only consistent gravitational theory within certain classes of anomaly-free $Sp(16)/\mathbb{Z}_2$ gauge theories may be a first tantalizing hint towards ``String Universality'' or ``String Lamppost Principle'' \cite{Kumar:2009us, Adams:2010zy, Kim:2019vuc, Kim:2019ths, Cvetic:2020kuw, Montero:2020icj, Hamada:2021bbz, Bedroya:2021fbu, Hamada:2023zol} --- that all consistent effective quantum gravity theories have a string theory realization --- also in the non-supersymmetric landscape.
With a better understanding of the charged defects developed in this work, the Sugimoto model may be a fruitful starting point for exploring these ideas.

\section*{Acknowledgements}
We are grateful to Ivano Basile, Miguel Montero, and Ingmar Saberi for sharing their insight and engaging in enlightening discussions with us.
Furthermore, we thank Markus Dierigl and H{\' e}ctor Parra de Freitas for valuable comments.

\appendix

\section{Characteristic classes}
\label{sec:Appendix A}

In this appendix we set the conventions and collect some useful formulae about characteristic classes that appear in anomaly computations of Section \ref{sec:anomaly_inflow}.

First, we follow a wide spread convention of changing the normalization of the curvature forms in comparison to \cite{Bilal:2008qx} as, 
\begin{equation}
    \frac{F}{2\pi}\longrightarrow F\,, \qquad \frac{R}{2\pi}\longrightarrow R\,.
\end{equation}
Following this choice, the Dirac genus reads:
\begin{equation}
\begin{split}
    \hat{A}(\mathcal{M}) & = 1+\frac{1}{2^2}\frac{1}{12}\tr R^2 + \frac{1}{2^4}\left[\frac{1}{360}\tr R^4 + \frac{1}{288}(\tr R^2)^2\right]+  \\
    & \, +\frac{1}{2^6}\left[ \frac{1}{5670}\tr R^6 + \frac{1}{4320}\tr R^4\tr R^2 + \frac{1}{10368}(\tr R^2)^3\right] + ... \\
    & = 1 - \frac{p_1(R)}{24} + \frac{7 p_1^2(R) - 4 p_2(R)}{5760} + \frac{-31p_1^3(R) + 44 p_1(R) p_2(R) - 16 p_3(R)}{967680} + ...\, ,
    \label{eq:Dirac genus}
\end{split}
\end{equation}
where the traces are in the fundamental representation, and we have used the Pontryagin classes
\begin{equation}
\begin{split}
     p_1(R) &= -\frac{1}{2}\tr{R^2}\,,\\
     p_2(R) &=\frac{1}{8}\left((\tr{R^2})^2-2\tr{R^4}\right)\, , \\
     p_3(R) &= -\frac{1}{48} \left( (\tr F^2)^3 - 6 \tr F^2 \tr F^4 + 8 \tr F^6 \right) \, .
\end{split}
\label{eq:Pontryagin classes}
\end{equation}
Under the decomposition ${\cal T}M|_{\Sigma} = {\cal T} \Sigma \oplus {\cal N}\Sigma$ of the tangent bundle on a submanifold $\Sigma$, these classes decompose as
\begin{equation}\label{eq:Pontryagin_decomposition}
\begin{split}
    p_1(R) &= p_1(N)+p_1(\tilde{R}) \, , \\
    p_2(R) &= p_2(\tilde{R})+p_2(N)+p_1(N)p_1(R)-p_1^2(N) \, .
\end{split}
\end{equation}
The Chern character is defined as
\begin{equation}
    \text{ch}_\mathcal{R}(F) = \tr_\mathcal{R} e^{iF} = \sum_{k=0}^\infty \frac{i^k}{k!} \tr_{\mathcal{R}}(F^k) \, .
    \label{eq:Chern character}
\end{equation}

For the anomaly computation on the defects, we also need the Chern characters for the associated bundles in the (co-)spinor representations of the transverse rotation symmetry. 
These can be computed easily following, e.g., \cite{Schellekens:1986xh}.
For the D1 (with transverse $\mathfrak{so}(8)_T$) the result is
\begin{equation}
    \text{ch}_{s/c}(N) = 8 +  p_1(N) + ...\, ,
    \label{eq:D1 Chern spin bundles}
\end{equation}
while in the case of the D5 (with transverse $\mathfrak{so}(4)_T$) we find
\begin{equation}
    \text{ch}_{{\bf 2}_\pm}(N) = 2 + \frac{p_1(N)\pm 2\chi(N)}{4} + \frac{p_1^2(N) +4p_2(N)\pm 4p_1(N) \, \chi(N)}{192} + ...\, ,
    \label{eq:D5 Chern spin bundles}
\end{equation}
and $\chi(N)$ is the Euler class.
Here we have neglected the higher-degree terms because they do not enter in the anomaly polynomials.

The Chern character of general representations can be deduced from the basic properties
\begin{align}
\begin{split}
    \text{ch}_{\mathcal{R}_1 \oplus \mathcal{R}_2} F & = \text{ch}_{\mathcal{R}_1} F + \text{ch}_{\mathcal{R}_2} F \, , \\
    \text{ch}_{\mathcal{R}_1 \otimes \mathcal{R}_2} F & = \text{ch}_{\mathcal{R}_1} F \wedge \text{ch}_{\mathcal{R}_2} F \, .
\end{split}
\end{align}
Some useful formulae for (anti-)symmetric tensor products are
\begin{align}
    \text{ch}_{\text{\tiny\yng(2)}} (F) & = \frac{1}{2}\left[(\text{ch} (F))^2 + \text{ch}(2F)\right] \, ,
    \label{eq:ch symm property} \\
    \text{ch}_{\text{\tiny\yng(1,1)}} (F) & = \frac{1}{2}\left[(\text{ch}( F)^2 - \text{ch}(2F)\right] \, ,
    \label{eq:ch anti-symm property}
\end{align}
where $\text{ch} \equiv \text{ch}_{\tiny\yng(1)}$.
For the gauge algebras $\mathfrak{g} = \mathfrak{so}, \mathfrak{sp}$ the vanishing of
\begin{equation}
    \tr_\mathcal{R} F^{2k+1}=0 \, ,
    \label{eq:Zero odd trace}
\end{equation}
in any representation implies that all anomaly polynomials that appear in this work only have terms of degrees 0 mod 4.

\subsection{Coefficients for rank-2 representations}
\label{sec:Appendix trace coefficients up to rank 2}
 Considering $\tr_\mathcal{R} F^2$, for the rank-2 symmetric representation we observe that
    \begin{equation}
            \tr_{\text{\tiny\yng(2)}} F^2 = -2 \left[\text{ch}_{\text{\tiny\yng(2)}}F\right]\Big|_{4} = -2 \frac{1}{2}\left[(\text{ch} F)^2 + \text{ch}(2F)\right]\Big|_{4}\,,
    \end{equation}
    where we used the property (\ref{eq:ch symm property}).
    Thus, we obtain, for the symmetric representation
    \begin{equation}
        \tr_{\text{\tiny\yng(2)}} F^2 = (n+2) \tr  F^2.
        \label{eq:trF^2 symm}
    \end{equation}
    An analogous calculation can be performed for the anti-symmetric representation
    \begin{equation}
    \left[\text{ch}_{\text{\tiny\yng(1,1)}} F\right]\Big|_{4} = -\frac{1}{2}\tr_{\text{\tiny\yng(1,1)}}F^2 =  \frac{1}{2}\left[(\text{ch} F)^2 - \text{ch}(2F)\right]\Big|_{4} ,
    \end{equation}
    leading us to
    \begin{equation}
        \tr_{\text{\tiny\yng(1,1)}} F^2 = (n-2) \tr  F^2.
        \label{eq:trF^2 anti-symm}
    \end{equation}
    In the same manner we will derive $\tr_\mathcal{R} F^4$ by noting that
    \begin{equation*}
        \frac{1}{24}\tr_\mathcal{R}  F^4 = [\text{ch}_\mathcal{R} F]|_8\,.
    \end{equation*}
    Indeed, we find:
    \begin{equation}
            \tr_{\text{\tiny\yng(2)}} F^4 = 24 \left[\text{ch}_{\text{\tiny\yng(2)}}F\right]\Big|_{8} = 24\frac{1}{2}\left[(\text{ch} F)^2 + \text{ch}(2F)\right]\Big|_{8},
    \end{equation}
    so
    \begin{equation}
        \tr_{\text{\tiny\yng(2)}} F^4 = (n+8) \tr  F^4 + 3(\tr  F^2)^2.
        \label{eq:trF^4 symm}
    \end{equation}
    Analogously, for the anti-symmetric we have
    \begin{equation}
        \tr_{\text{\tiny\yng(1,1)}} F^4 = (n-8) \tr  F^4 +3(\tr  F^2)^2.
        \label{eq:trF^4 anti-symm}
    \end{equation}
    In order to understand the bulk anomaly computations, it is useful to compute the coefficients for $\tr_{\mathcal{R}}  F^6$
    \begin{equation}
        \begin{split}
            \tr_{{\text{\tiny\yng(1,1)}}} F^6 =& -720 \left[\text{ch}_{\text{\tiny\yng(1,1)}}(F)\right]\Big|_{12} = -\frac{720}{2}\left[(\text{ch} F)^2 - \text{ch}(2F)\right] \Big|_{12},
        \end{split}
    \end{equation}
    resulting in:
    \begin{equation}
        \tr_{{\text{\tiny\yng(1,1)}}} F^6  = (n-32) \tr F^6 + 15 \tr F^2\tr F^4.
        \label{eq:trF^6 anti-symm}
    \end{equation}
    For the symmetric representation, we have
    \begin{equation}
        \tr_{{\text{\tiny\yng(2)}}} F^6  = (n+32) \tr F^6 + 15 \tr F^2\tr F^4.
        \label{eq:trF^6 symm}
    \end{equation}

\section{Trace coefficients for higher rank \texorpdfstring{$\mathfrak{sp}(16)$}{sp(16)} representations}
\label{sec:Appendix higher rank trace coefficient}

For sake of completeness, it is crucial to illustrate the methodology that we employed for computing the traces of the higher rank degrees of freedom employed in section \ref{sec:Higher rank particles}. Following the arguments in \cite{Basile:2023zng}, we consider the graded Chern character of the exterior algebra $\Lambda(V) = \oplus_n \Lambda^n(V)$ of a bundle $V$
    \begin{equation}
        \text{ch}(\Lambda(V)) = \sum_n t^n \text{ch}(\Lambda^n(V)),
        \label{eq:Graded Chern character}
    \end{equation}
    the same can be done with the symmetric algebra $S(V) = \oplus_n S^n(V)$ if we want to compute the traces related to symmetric representation. Both these algebras can be decomposed as a product of line bundles ones, $\Lambda(V) = \otimes_i \Lambda(\mathcal{L}_i)$ and  $S(V) = \otimes_i S(\mathcal{L}_i)$. The related graded Chern characters are
    \begin{equation}
    \begin{split}
        \text{ch}(\Lambda(\mathcal{L})) &= 1 + t e^{c_1(\mathcal{L})}\,, \\
        \text{ch}(S(\mathcal{L})) &= 1 + t e^{c_1(\mathcal{L})} + ... = \frac{1}{1-te^{c_1(\mathcal{L})}}\,.
    \end{split}       
    \end{equation}
    Thanks to the multiplicative property of the Chern characters we can state that:
    \begin{equation}
    \begin{split}
        \text{ch}(\Lambda(V)) & = \prod_i \left( 1 + t e^{c_1(\mathcal{L}_i)}\right) = \det(1 + te^{F}) =e^{\tr \log(1+te^{F})}\,, \\
        \text{ch}(S(V)) &= \prod_i \frac{1}{1-te^{c_1(\mathcal{L}_i)}} =  \frac{1}{\det(1-te^{F})} = e^{-\tr \log(1-te^{F})}\,.
        \label{eq:Graded Chern character final form}
    \end{split}
    \end{equation}
    Equating the correct terms in the expansions of (\ref{eq:Graded Chern character}) and (\ref{eq:Graded Chern character final form}), is possible to compute the relations for the traces of the representations employed in this work. They are presented in the following.
\paragraph{Rank 4 symmetric}
\begin{equation}
    \tr_{\text{\tiny{\yng(4)}}}\textbf{1} = \frac{n(n+1)(n+2)(n+3)}{24}
    \label{eq:Rk4 sy tr1}
\end{equation}\\
\begin{equation}
    \tr_{\text{\tiny{\yng(4)}}} F^2 = \frac{(n+2)(n+3)(n+4)}{6} \tr F^2
    \label{eq:Rk4 sy trF^2}
\end{equation}\\
\begin{equation}
    \tr_{\text{\tiny{\yng(4)}}} F^4 = \frac{(n+4)(n^2 + 23 n + 96)}{6} \tr F^4 + \frac{3(n+4)(n+5)}{2} (\tr F^2)^2
    \label{eq:Rk4 sy trF^4}
\end{equation}\\
\begin{equation}
\begin{split}
    \tr_{\text{\tiny{\yng(4)}}} F^6 =& \frac{(n^3+99 n^2 + 1556 n+6144)}{6} \tr F^6 +\\
    &+\frac{45 (n^2+ 21 n + 92 )}{6} \tr F^4 \tr F^2 + \frac{90(n+6)}{6} (\tr F^2)^3
    \label{eq:Rk4 sy trF^6}
\end{split}
\end{equation}

\paragraph{Rank 4 antisymmetric}
\begin{equation}
    \tr_{\text{\tiny{\yng(1,1,1,1)}}}\textbf{1} = \frac{n(n-1)(n-2)(n-3)}{24}
    \label{eq:Rk4 asy tr1}
\end{equation}\\
\begin{equation}
    \tr_{\text{\tiny{\yng(1,1,1,1)}}} F^2 = \frac{(n-2)(n-3)(n-4)}{6} \tr F^2
    \label{eq:Rk4 asy trF^2}
\end{equation}\\
\begin{equation}
    \tr_{\text{\tiny{\yng(1,1,1,1)}}} F^4 = -\frac{27n^2 - 188 n + 384}{6} \tr F^4 + \frac{3(n-4)(n-5)}{2} (\tr F^2)^2
    \label{eq:Rk4 asy trF^4}
\end{equation}\\
\begin{equation}
\begin{split}
    \tr_{\text{\tiny{\yng(1,1,1,1)}}} F^6 =& \frac{(n^3 - 99 n^2 + 1556 n  -6144 )}{6} \tr F^6 + \frac{45 (n^2 - 21 n + 92 )}{6} \tr F^4 \tr F^2 +\\
    &+ \frac{90(n-6)}{6} (\tr F^2)^3
    \label{eq:Rk4 asy trF^6}
\end{split}
\end{equation}

\paragraph{Rank 6 symmetric}
\begin{equation}
    \tr_{\text{\tiny{\yng(6)}}}\textbf{1} = \frac{n(n+1)(n+2)(n+3)(n+4)(n+5)}{720}
    \label{eq:Rk6 sy tr1}
\end{equation}\\
\begin{equation}
    \tr_{\text{\tiny{\yng(6)}}} F^2 = \frac{(n+2)(n+3)(n+4)(n+5)(n+6)}{120} \tr F^2
    \label{eq:Rk6 sy trF^2}
\end{equation}\\
\begin{equation}
\begin{split}
    \tr_{\text{\tiny{\yng(6)}}} F^4 =& \frac{(n+4)(n+5)(n+6)(n+8)(n+27)}{120} \tr F^4 +\\
    &+ \frac{(n+4)(n+5)(n+6)(n+7))}{8} (\tr F^2)^2
    \label{eq:Rk6 sy trF^4}
\end{split}
\end{equation}\\
\begin{equation}
\begin{split}
    \tr_{\text{\tiny{\yng(6)}}} F^6 = & \frac{(n+6)(n^4 + 164n^3 + 4871 n^2 + 48604 n + 155520)}{6} \tr F^6 +\\
    & +\frac{5(n+6)(n+7)(n^2 +33 n + 212 )}{8} \tr F^4 \tr F^2 +\\
    &+ \frac{5(n+6)(n+7)(n+8)}{2} (\tr F^2)^3
    \label{eq:Rk6 sy trF^6}
\end{split}
\end{equation}

\bibliographystyle{JHEP}
\bibliography{bibliography.bib}

\end{document}